%% file: pre-print.tex
\documentclass{amsart}

\usepackage[nolist]{acronym}
\usepackage{float}
\usepackage[justification=centering]{subcaption}
\usepackage{amsmath}
\usepackage{graphicx}
\usepackage{xcolor}
\usepackage{amsfonts}
\usepackage{multirow}
\usepackage{graphics}
\usepackage{tabularx}
\usepackage{orcidlink}

\usepackage{booktabs}
\usepackage{tikz}

\usepackage{hyperref}
\hypersetup{colorlinks=true,linkcolor=blue,filecolor=magenta,urlcolor=blue,citecolor=red}

\usepackage[square,numbers]{natbib}

\theoremstyle{definition}

\theoremstyle{remark}

\numberwithin{equation}{section}

\begin{document}

\title[Agent-based Model of Initial Token Allocations]{Agent-based Model of Initial Token Allocations: Evaluating Wealth Concentration in Fair Launches}


%    author one information
\author{}
\address{}
\curraddr{}
\email{}
\thanks{}

\author[Delgado]{Joaquin Delgado Fernandez \orcidlink{0000-0003-1326-6134}}
%\authornote{Corresponding author.}
\address{SnT - Interdisciplinary Centre for Security, Reliability and Trust, University of Luxembourg, 29 Av. John F. Kennedy, Luxembourg}
\email{joaquin.delgadofernandez@uni.lu}

\author[Barbereau]{Tom Barbereau \orcidlink{0000-0002-8554-0991}}
\address{SnT - Interdisciplinary Centre for Security, Reliability and Trust, University of Luxembourg, 29 Av. John F. Kennedy, Luxembourg}
\email{tom.barbereau@uni.lu}

\author[Papageorgiou]{Orestis Papageorgiou \orcidlink{0000-0002-2245-5079}}
\address{SnT - Interdisciplinary Centre for Security, Reliability and Trust, University of Luxembourg, 29 Av. John F. Kennedy, Luxembourg}
\email{orestis.papageorgiou@uni.lu}

%    \subjclass is required.
%\subjclass[]{ }

%\date{2022}

\dedicatory{}

%    Abstract is required.
\begin{abstract}
 With advancements in distributed ledger technologies and smart contracts, tokenized voting rights gained prominence within Decentralized Finance (DeFi). Voting rights tokens (aka. governance tokens) are fungible tokens that grant individual holders the right to vote upon the fate of a project. The motivation behind these tokens is to achieve decentral control. Because the initial allocations of these tokens is often un-democratic, the DeFi project Yearn Finance experimented with a fair launch allocation where no tokens are pre-mined and all participants have an equal opportunity to receive them. Regardless, research on voting rights tokens highlights the formation of oligarchies over time. The hypothesis is that the tokens' tradability is the cause of concentration. To examine this proposition, this paper uses an Agent-based Model to simulate and analyze the concentration of voting rights tokens post fair launch under different trading modalities. It serves to examine three distinct token allocation scenarios considered as fair. The results show that regardless of the allocation, concentration persistently occurs. It confirms the hypothesis that the disease is endogenous: the cause of concentration is the tokens tradablility. The findings inform theoretical understandings and practical implications for on-chain governance mediated by tokens.
\end{abstract}

\maketitle
\raggedbottom

\input{sections/01_introduction}

\input{sections/02_literature}

\input{sections/03_analysis}

\input{sections/04_1_model}

\input{sections/04_2_metrics}

\input{sections/04_3_calibration}

\input{sections/05_results}
\input{sections/05_verification}

\input{sections/06_discussion}

\input{sections/07_conclusion}

\input{sections/08_aknowledgments}

\input{sections/99_acronyms}


\newpage

\bibliographystyle{abbrvnat}

\bibliography{references}

\end{document}

%% file: sections/01_introduction.tex
\section{Introduction}
\label{sec:intro}

%The inception of Bitcoin \citep{nakamoto_bitcoin_2008} fostered a novel movement of developers that began developing standards promoting ideals of decentralization beyond technical terms. Inherently, \textit{blockchain cultures} at large view decentralization in an ambiguous mix between political, economic, and organizational ideals \citep{schneider_decentralization_2019}. While 'decentralization' materializes differently in respective protocols, the Ethereum protocol stands out given its relatively high degrees of interoperability and programmability. From a technical standpoint, the disruptive innovation brought by it are \textit{smart contracts} – programmable and self-executing code deployed as part of \ac{DLT} or blockchain \citep{beck2018governance}. Furthermore, it includes \acp{ERC} – specific types of smart contracts and standards for the creation of fungible and non-fungible tokens. Ethereum was developed alongside a Turing-complete language that enables developers to leverage smart contracts and build \acp{dApp} and \acp{DAO} that strive for the said ideals. 

%Developments in Ethereum fostered the \ac{ICO} boom of 2017 that saw a panoply of projects flooding the market –- few of which survived the hype curve \citep{fridgen2018_ICO}.

The digital representation of value and ownership, in the form of fungible and non-fungible tokens respectively, provides the basis for the \textit{token economy}. Unlike traditional economies, the token economy does not rely on trusted third parties to verify transactions \citep{sunyaev2021token}, instead \ac{DLT} and smart contracts ensure integrity in a pseudonymous peer-to-peer network of interactions \citep{beck2018governance,zhang2019security}. 
%By leveraging \ac{DLT} and more specifically, \textit{smart contracts}, developments in the Ethereum community introduced financial services and products on platforms designed to operate without the involvement of trusted-third parties. 
Advancements in the token economy with a focus on financial services and products materialized under the heading of \ac{DeFi} \citep{zetzsche_decentralized_2020}. 

%While in public protocols 'decentralization' in technical terms is achieved, in political terms it is often contentious \citep{schneider_decentralization_2019}. This was observed in both Bitcoin and Ethereum. Over the decision to increase the size of Bitcoin blocks, community debates led to an outright ''civil war'' \citep[8]{de_filippi_invisible_2016} that pitted parties against each other over socio-political motives. The core developers, who act as gatekeepers to protocol changes, eventually took an autocratic decision against the increase -- a behavior pejoratively described as ''senatorial'' \citep{parkin_senatorial_2019}. The Ethereum protocol too faced its share of controversy: following an exploited bug in an Ethereum-based \ac{DAO}, \textit{The DAO}, the Ethereum Foundation's leaders decided to irreversibly fork the ledger \citep{dupont_chapter_2017}. Against this backdrop, \citet[19]{penzo2021hard} denote how in these public permissionless protocols, communities in power resort to informal adjudications while ``immune [...] from state scrutiny''.

%An alternative way to achieve decentral control and political decentralization, is by deploying 
\ac{DLT} enables people to coordinate themselves \textit{on-chain}, that is, transactions and interactions are ``mediated by a set of self-executing rules [i.e., smart contracts] deployed on a public blockchain'' independently from central control \citep[p. 1]{hassan2021decentralized}.  
To achieve decentral control in \ac{DeFi}, , developers created and allocated so-called \textit{voting rights tokens} – (fungible) tokens that stipulate voting entitlements to initiate changes to a public blockchain platform \citep{oliveira2018token}. 
%These tokens are allocated among (pseudonymous) community members with the ambition to distribute and share decision-making power. 
%These type of tokens were firstly introduced in the \ac{DeFi} space as means to form political consensus in communities of \acp{DAO} -- a manifestation of on-chain governance. 
An example is the \ac{DEX} Uniswap, a platform which uses smart contracts to automate the exchange of fungible tokens of the Ethereum protocol. Its voting rights token, UNI, allows holders to cast votes and decide upon the use of resources stored in a treasury (defined in a smart contract). 
%While other \acp{DEX} are SushiSwap and PancakeSwap, Aave and Compound offer decentralized lending services, Synthetix and UMA offer decentralized derivatives, and Nexus Mutual provides a decentralized insurance model \citep{amler2021defi}. All of these include a respective \ac{DAO} plus voting rights tokens and belong to the Bloomberg Galaxy \ac{DeFi} Index (Total value locked approximately \$43 billion in late April 2022). 
The distribution of voting rights tokens, however, is considered as controversial given that the initial allocation often favors a minority of insiders (e.g., developers, investors, etc.). Tokens allocated to insiders are common within \acp{ICO} \citep{bourveau2022role}, hence a differentiation between initial allocations that favor insiders (labeled, ``private'') or those that do not (labeled, ``public'') \citep[p. 10]{fridgen2018_ICO}. History repeats itself as insider allocations of voting rights tokens are common in \ac{DeFi} \citep{barbereau_decentralised_2022}.

%Recent descriptive research, however, demonstrated how the success of the platforms' voting rights tokens falls short in achieving political decentralization. Indeed, in an analysis of nine Ethereum-based platforms and the distribution over time of their voting rights tokens, \citet{barbereau2022defi:HICSS} showcase that eventually they lead to a form of oligarchy – one whereby the few hold the majority of these tokens. 
One outlier to insider allocations is Yearn Finance. Its core developer, Andre Cronje, denoted that in \ac{DeFi} voting rights tokens are majoritarily allocated to a community of ``friends and family'' \citep{cronje_fair_2021} -- impossibly leading to decentral control. As solution, he used a type of initial token allocation for the voting rights tokens of Yearn Finance (YFI) –- the \textit{fair launch} -- whereby all community members have an equal opportunity to receive a portion of the initial supply \citep{cronje_podcast_2020}.  Though in theory this allocation strategy achieves equity through principles of fairness \citep{rawls1991justice}, reality looks dire in the long run: \citet{barbereau2022defi:HICSS} demonstrate how, as with all other voting rights tokens in \ac{DeFi}, for YFI concentration of wealth and voting power persists. 

Because holders rarely use these to cast votes, \citet{barbereau_decentralised_2022} denote a common theme and propose a purposefully \textit{descriptive} theory of voting rights tokens as justification for concentration: they are \textit{tradable} assets on cryptocurrency markets. This description may not seem surprising against the consideration that wealth in the token economy is concentrated (c.f. concetration in Bitcoin and Ethereum \citep{gochhayat_measuring_2020}), and so are capital markets more broadly \citep{piketty_capital_2014}. Indeed, the common feature of tradability appears to justify,  on an intuitive level, the expectations that ``wealth trickles up in free-market economies'' \citep{boghosian2019inequality}.

The hypothesis whether the experiment of Andre Cronje's fair launch was inherently doomed to fail given that the underlying tokens are tradable remains untested. 
%The extent to which trading affects concentration post initial allocation presents a research opportunity. 
%In parallel, within the cross-disciplinary topic of \textit{tokenomics} \citep{oliveira2018token}, initial token allocations also present a research opportunity. 
%\subsection{Research Questions \& Contributions}
Taking the principle of a 'fair' launch (n.b., all tokens are allocated fairly at initialization), one can consider weather alternative allocations are successful in achieving decentral control in the long run. Correspondingly, to challenge this hypothesis, this article addresses the following research questions: 

\begin{itemize}
    \item[\textbf{RQ1}:] Does trading behavior affect voting rights token distributions over time?
    \item[\textbf{RQ2}:] Do alternative, 'fair launch' token allocations affect voting rights token distributions over time?
\end{itemize}

Provided the financial context in which these novel governance structures are deployed, our research topic -- 'fair launch' token allocations -- is of importance to \ac{IS} and requires multidisciplinary perspectives \citep{treiblmaier2021_whatsnext}. Hence, we lean on previous theory on governance of public protocols and token design, and use quantitative methods rooted in simulation. Guided by an ambition ``for discovery and explanation'' \citep[p. 516]{beese2019simulation}, we adopt \ac{ABM} to simulate the trade and eventual distribution of voting rights tokens post three distinct `fair launch' allocations (n.b., scenarios denoted $S_n$). The developed model is going from ``real world to simulation world'' \citep[p. 516]{beese2019simulation}, an approach that is particularly suitable to the exploration of novel phenomena -- here, the fair launch. Within \ac{IS}, the utility of \ac{ABM} for the study of phenomena with ``nonlinear behavior'' \citep[p. 158]{haki2020evolution} is well-recognized. At large, the discipline is receptive of contributions emerging from simulation research \citep{davis2007developing,zhang2014rethinking,beese2019simulation}, justifying the use of \ac{ABM} for this study.

%\citet[p. 482]{davis2007developing} argue that the assumptions underlying the \ac{ABM} ought to be made on the basis of ''simple theory''. On the basis of \citet{cocco2017_abm_crypto} and \citet{rocsu2021evolution}, we assume that each agent represents a trader (and consequently, a potential voter) who actively partakes in an artificially defined cryptocurrency market. The trading behavior of agents within that environment is informed by individual wealth and market conditions. The market rules follow principles of a clearing house\citep{mendelson1982market}. 
%\footnote{Given the computational intensity of a price clearing and the focus on token concentration, we set the price of the orders to follow Yearn Finance's.} 
%Further, the model is informed by real world data extracted from the Ethereum ledger about Yearn Finance. The created fair launch token allocation scenarios are loosely based on fairness principles discussed in political philosophy \citep{mill_1864,rawls1971theories} notably, social liberalism (for $S_0$), egalitarianism (for $S_1$), and Darwinism (for $S_2$). 

\citet[p. 482]{davis2007developing} advise to ground the model within ``simple theory''; theory, that provides the ``basic concepts and process that describe a phenomenon'' \citep[p. 506]{beese2019simulation}. Here, we focus on governance of public protocols (Section \ref{sec:lit}). To establish further ``epistemic credibility in the simulation model'' \citep[p. 517]{beese2019simulation}, aside from theory (deductive approach) we use empirical data (inductive approach) from Yearn Finance (Section \ref{sec:data_prep}). 

The development of our model (Section \ref{sec:model}) is informed by the artificial cryptocurrency markets designed in \citet{cocco2017_abm_crypto} and \citet{rocsu2021evolution}. Therein, agents represent traders that are endowed with an amount of fiat currency and seek to acquire the (artificially created) voting rights tokens (TKNs). The market rules are loosely based on understandings of clearing houses \citep{mendelson1982market}. To investigate RQ2, we developed two alternative scenarios to Andre Cronje's fair launch scenario ($S_0$) which respectively consider fairness in egalitarian terms ($S_1$) and 'at random' ($S_2$). The principles underlying these allocations build on political philosophy \citep{mill_1864,rawls1971theories}. 
We measure concentration in terms of the Gini Coefficient \citep{1912vamu.book.....G} and the Shannon Entropy \citep{shannon_mathematical_1948}. With our work we make the following contributions:

\begin{itemize}
    \item An agent-based model for the analysis of token distributions under various market conditions reflective of trading.
    \item Simulation results showing how over time, regardless of initial token allocation, concentration is imminent.
    \item Extended understandings on tokenomics to formerly include token allocations as part of governance parameters.
\end{itemize}

%\reviewer{Writing-wise, the introduction section misses a clear summary of contributions.}\tbnote{Done above. Please complement}

%% file: sections/02_literature.tex
\section{Related Work}
\label{sec:lit}

%%%%%%%%%%%%%%%%%%%%%%%%%%%%%%%%%%%%%%%%%%%%%%%%%%%%%%%%%%%
\subsection{Governance in Public-protocols}

Bitcoin \citep{nakamoto_bitcoin_2008} led to a burgeoning movement of developers that saw decentralization beyond technical terms; not least, as an ambiguous mix of political, economic, and organizational ideals \citep{schneider_decentralization_2019} that materialized in the emergence of alternative protocols. Commonly the next generation of \ac{DLT} includes \textit{smart contracts} and the possibility to deploy tokens \citep{beck2018governance,zhang2019security}. 
%Scholars study tokens under the heading of \textit{tokenomics} -- a subdomain of cross-disciplinary research on \ac{DLT} \citep{oliveira2018token}. 
%\reviewer{ERC20 and ERC721 are specific token standards. ERC in general is just an organizational structure to come up with community wide standards. It is also not specific to Ethereum Bitcoin similarly has  BIPs. In think, the whole paragraph could be shortened and made more concise. Much of the details and buzzwords such as "turing complete language", "Dapps" are just not necessary.}\tbnote{i think i addressed this comment with the current changes}
While in public protocols, 'decentralization' in technical terms is achieved, in political terms, it is often contentious \citep{schneider_decentralization_2019}. Over the decision to increase the size of Bitcoin blocks, community debates led to an outright "civil war" \citep[p. 8]{de_filippi_invisible_2016} that pitted parties against each other over socio-political motives. The core developers, who act as gatekeepers to protocol changes, eventually took an autocratic decision against that increase -- a behavior pejoratively described as "senatorial" \citep{parkin_senatorial_2019}. The Ethereum protocol, too, faced its share of controversy: following an exploited smart contract bug the Ethereum Foundation's leaders decided to irreversibly fork the ledger \citep{dupont_chapter_2017}. 
Against this backdrop, \citet[19]{penzo2021hard} denote how in public permissionless protocols, governing communities resort to informal adjudications, typically ``immune [...] from state scrutiny".

%Governance in \ac{DLT} systems was studied in various disciplines, including management and economics, as well as social science and law \citep{rossi2019blockchain,beck2018governance}. 
Consequently, scholars distinguish between \textit{on-chain} and \textit{off-chain} governance. The former refers to rules that enforce the 'code-is-law' dictum \citep{wright_decentralized_2015}; in other words, using smart contracts to define governance mechanisms and structures (“now the code runs itself”) \citep{reijers_blockchain_2018}. The informal resolution mechanisms in Bitcoin and Ethereum, however, are examples that demonstrate the shortcomings of the ditcum and shed light on ulterior power structures \citep{wright_decentralized_2015,de_filippi_invisible_2016,dupont_chapter_2017}. Off-chain governance refers to the formalization of control via the intermediary of endogenous (e.g., through the foundation of institutions such as consortia, cooperatives, etc.) or exogenous (e.g., national laws, regulations, standards, etc.) structures \citep{reijers_blockchain_2018,ziolkowski2020decision}. 
%The dictum alludes to the unique authority of code to regulate conduct (and thereby governance) within a system -– a notion that transcended into early blockchain cultures drawn towards decentralized and increasingly autonomous systems \citep{schneider_decentralization_2019}. The DAO was a first manifestation of systems fully governed \textit{on-chain} \citep{dupont_chapter_2017}.

%Although a failed project, The DAO was a first manifestation of systems fully governed \textit{on-chain} \citep{dupont_chapter_2017}. Bitcoin, in turn, is a hybrid of sorts. Its numerous forks are a manifestation of on-chain governance \citep{andersen2019self}. The (senatorial) initialization thereof by the Bitcoin core developers following the block size controversy, for instance, is seen in the perspective of off-chain governance \citep{de_filippi_invisible_2016,parkin_senatorial_2019}.

%Though the code-is-law dictum refers to the reduction of transaction costs in favor of 'trustless' and 'autonomous' environments, paradoxically, it also leaves space for the establishment of institutional structures of power, particularly in public permissionless systems \citep{wright_decentralized_2015,pazaitis_blockchain_2017,vergne2020decentralized}. The story of The DAO hack and Bitcoin's block size controversy are often cited examples that highlight the shortcomings of the code-is-law dictum and the need for governance beyond the technical layer. 

%%%%%%%%%%%%%%%%%%%%%%%%%%%%%%%%%%%%%%%%%%%%%%%%%%%%%%%%%
\subsection{Governance in Decentralized Finance}

Early research on \ac{DLT} praised the ``ability to cut out the middleman'' \citep{underwood_beyondBTC_2016} in financial applications. With \acf{DeFi}, expectations became reality in early 2020. The certainly powerful premise to interact without intermediaries to lend or stake tokens (in exchange for high interest and other rewards) generate a significant buzz. Uniswap and SushiSwap are \acp{DEX}; Aave and Compound offer decentralized lending services; Synthetix and UMA offer decentralized derivatives; and Nexus Mutual provides a decentralized insurance model \citep{amler2021defi}. All of these include a respective \ac{DAO} plus voting rights tokens and belong to the Bloomberg Galaxy \ac{DeFi} Index (Total value locked approximately \$43 billion in late April 2022). Rightfully, \citet{subramanian_DLT_marketplace_2018} denoted how decentralized electronic marketplaces can, ``if successful, complement and rival'' traditional ones.

Within \ac{DeFi}, beyond improvements made to the financial value chain \citep[]{schar2021decentralized}, experiments were made at implementing governance structures fully on-chain; most notably, by embedding voting rights into tokens. These tokens grant holders the capacity to cast votes on proposals. While the features of these tokens are contextual to the platform, the majority of these follow the fungible token standard ERC-20. Like most cryptocurrencies, they are tradable on regular and decentralized exchanges \citep{barbereau2022defi:HICSS}. By nature, the study of these tokens is at the intersection of research on blockchain governance and \textit{tokenomics} -- a subdomain of cross-disciplinary research on \ac{DLT}.

%The distribution of voting rights tokens, ubiquitous in the field of \ac{DeFi}, is subject to ample controversies. 
\citet[p.8]{oliveira2018token} define "Governance Parameters" of tokens as those parameters that "relate to what [it] effectively represents and how this connects to the way the platform is governed and managed". The authors introduce three parameters (Table \ref{tab:token_classification}): (1) "Representation" (the type of asset represented by a token), (2) "Supply" (the way tokens are distributed), and (3) "Incentive system" (the way a token exerts influence over the network and/or its holder). Our scope is on token allocations and distributions, hence of primary relevance being the "Supply" parameter. \citet[p. 9]{oliveira2018token} note how "Supply" strategies can either be on a one-time basis ("fixed") or following increments ("schedule-based"). Tokens can also be "pre-mined" (or "pre-sold" \citep{fridgen2018_ICO}), that is a portion of the tokens is created and distributed before the official launch date.

\begin{table}[ht!]
\renewcommand{\arraystretch}{2}
\caption{Excerpt of the Token Classification proposed in \citet{oliveira2018token}. \label{tab:token_classification}}

\centering
\resizebox{\columnwidth}{!}{%
\begin{tabular}{|l|l|c|llllllllllll} 
\hline
\multirow{3}{*}{\vspace{-0.8cm}\textbf{Governance Parameters}} & \multicolumn{2}{c|}{\textbf{Representation}}            & \multicolumn{4}{>{\centering\arraybackslash}p{4cm}|}{Digital}           & \multicolumn{4}{>{\centering\arraybackslash}p{4cm}|}{Physical}                             & \multicolumn{4}{>{\centering\arraybackslash}p{4cm}|}{Legal}                                                            \\ 
\cline{2-15}
                                       & \multicolumn{2}{c|}{\textbf{Supply}} &           \multicolumn{3}{>{\centering\arraybackslash}m{3cm}|}{Schedule-based} & \multicolumn{3}{>{\centering\arraybackslash}m{3cm}|}{Pre-mined, scheduled distribution} & \multicolumn{3}{>{\centering\arraybackslash}m{3cm}|}{Pre-mined, one-off distribution} & \multicolumn{3}{>{\centering\arraybackslash}m{3cm}|}{Discretionary}   \\ 
\cline{2-15}
                                       & \multicolumn{2}{c|}{\textbf{Incentive system}}          & \multicolumn{3}{c|}{Enter Platform} & \multicolumn{3}{c|}{Use Platform}                      & \multicolumn{3}{c|}{Stay Long-Term}                  & \multicolumn{3}{c|}{Leave Platform}  \\ 
\hline
                       
\end{tabular}%

}
\end{table}

For the supply of voting rights tokens, whose fair deployment is motivated by a normative ambition of political decentralization, the story is more ambiguous. Uniswap developers pre-mined a part of all voting rights tokens (UNI) and allocated some to a group of insiders. Among others, the \ac{DeFi} projects SushiSwap (SUSHI) and MakerDAO (MKR) followed similar paths, opting for an allocation that favored insiders. Over time, in all of these cases, wealth concentration was eminent \citep{barbereau_decentralised_2022}. 

Concentration of wealth and power is inherent to human societies and economic systems \citep{piketty_capital_2014}. \citet{pareto1964cours} exposed the land concentration in the Italian \textit{novecento}; subsequently lending his name to the Pareto Principle. Financial markets are no exception to the principle
%: while in the 1980s the top 10 banks held 25\% of total wealth, in 2008 (30 years later) the top 10 banks held the 50\% of the industry assets 
\citep{mester2007some}. 
%\tbnote{maybe you can use this as transition} Given the tradability of these voting rights tokens, the trends of concentration over time are consistent with the expectation of \citet{boghosian2019inequality} that "wealth trickles up in free-market economies". This was previously demonstrated for Bitcoin and Ethereum \citep{gochhayat_measuring_2020}.
In cryptocurrency markets the same occurs: both Bitcoin and Ethereum have centralized token distributions, and the trend appears to only increase \citep{gochhayat_measuring_2020}. This is true even in the case of Proof-of-Stake cryptocurrencies where the project's security is closely tied to the level of dispersion \citep{rocsu2021evolution}. When it comes to voting rights tokens, \citet{barbereau2022defi:HICSS} identified that the level of concentration among voting rights tokens is even higher, highlighting cases where a handful of people hold more than 50\% of all tokens. 

Andre Cronje's project Yearn Finance (YFI) sought to eliminate favoritism and insider allocations \citep{cronje_podcast_2020}. By opting for the first, fixed supply strategy, YFI were \textit{not} allocated to a minority of insiders. The implemented \textit{fair launch} followed the principle of 'fair equality of opportunity' \citep{rawls1991justice}; effectively, the idea that each user has the same opportunity to obtain YFIs. Despite its failure to achieve an equitable distribution over time \citep{barbereau2022defi:HICSS}, at least in theory, a fixed initial token allocation that is 'fair' would help achieve ambitions of decentral control.

%%%%%%%%%%%%%%%%%%%%%%%%%%%%%%%%%%%%%%%%%%%%%%%%%%
\subsection{Agent-based modeling}
%\tbnote{we have to see if this paragraph moves elsewhere} 

%To evaluate the phenomenon of initial token allocations and concentration of tokens we designed an \acf{ABM}. We do so on the basis of "simple theory" \citep[p. 482]{davis2007developing}.
%Prior to describing the data preparation and model design in detail, subsequently we discuss a humble fraction of the literature on \ac{ABM}. 
\ac{ABM} is a computational method used to simulate the actions and/or interactions of autonomous agents in order to understand how systems behave and what determines outcomes \citep{macal2016everything}. As analytical method, applications of \ac{ABM} are found in a variety of disciplines from energy and pathology to risk management and finance. 
%Most recently, the application of \ac{ABM} in epidemiology coincided with the COVID-19 pandemic where it was used to model the spread of the SARS-CoV-2 virus among populations \citep{rockett2020revealing} and simulate transmission risks in hospitals \citep{cuevas2020agent}. 
%In economics, then, despite significant work in computational finance and economics more broadly (e.g., \citep{lebaron2006agent, fagiolo2019validation}), \citet{farmer2009economy} argue for an increased integration of \ac{ABM} to develop economic models and inform policy. 
%Though numerous definitions were proposed in literature, we consider an \ac{ABM} in which "the individual agents have internal behaviors that allow them to be autonomous, able to sense whatever condition occurs within the model at any time, and to act on the appropriate behavior in response" \citep[p. 149]{macal2016everything}. 
Scholars acknowledged the value and potential of \ac{ABM} for \ac{IS} research given its methodological and analytical versatility to investigate systems whose "emergent properties unfold over time" \citep[p. 158]{haki2020evolution} and value to generate theory \citep{davis2007developing}. The literature review of \citet{beese2019simulation} illustrates the breadth of \ac{ABM} applications in \ac{IS} -- citing the potential for researchers to embed theory in the exploration of complex phenomena.

%\citet{schalowski_long-term_2019} studied the effects of market fluctuations on platform diffusion. 
%\citet{torres_pena_crafting_2019} focused on (service) platforms and used \ac{ABM} to propose an analytical model for the study of platform dynamics. \citet{onuchowska_using_2019} evaluated the effect of different social media policies for the prevention of malicious behavior in humans and bots. Blockchain systems, then, were studied by \citet{hulsemann_walk_2019} who used \ac{ABM} to evaluate the incentive structures of different token designs for \acp{dApp}. 

For the study of cryptocurrencies and blockchain-based systems at large, scholars applied \ac{ABM} in several contexts. \citet{bornholdt2014bitcoins} proposed a model to study the emergence of cryptocurrencies vis-à-vis Bitcoin -- considering factors such as trading, mining of new coins, and agent-to-agent interactions. Their findings show that Bitcoin may be interchangeable with cryptocurrencies of similar characteristics. \citet{cocco2017_abm_crypto} built an artificial cryptocurrency marketplace based on an order book simulation of the Bitcoin market where agents trade autonomously. Their model is able to reproduce real price formations and market volatility; hence, our adaptation of it in this work. \citet{rocsu2021evolution} propose an environment to model the behavior of investors/agents in a Proof-of-Stake (PoS) based blockchain of cryptocurrency issuance. They denote, contrary to expectations, that agents seek to stabilize their portfolio instead of accumulating more wealth.

%% file: sections/03_analysis.tex
\section{Data preparation}
\label{sec:data_prep}

\begin{figure}[ht!]
    \centering
    \makebox[0pt]{
    \includegraphics[width=\textwidth]{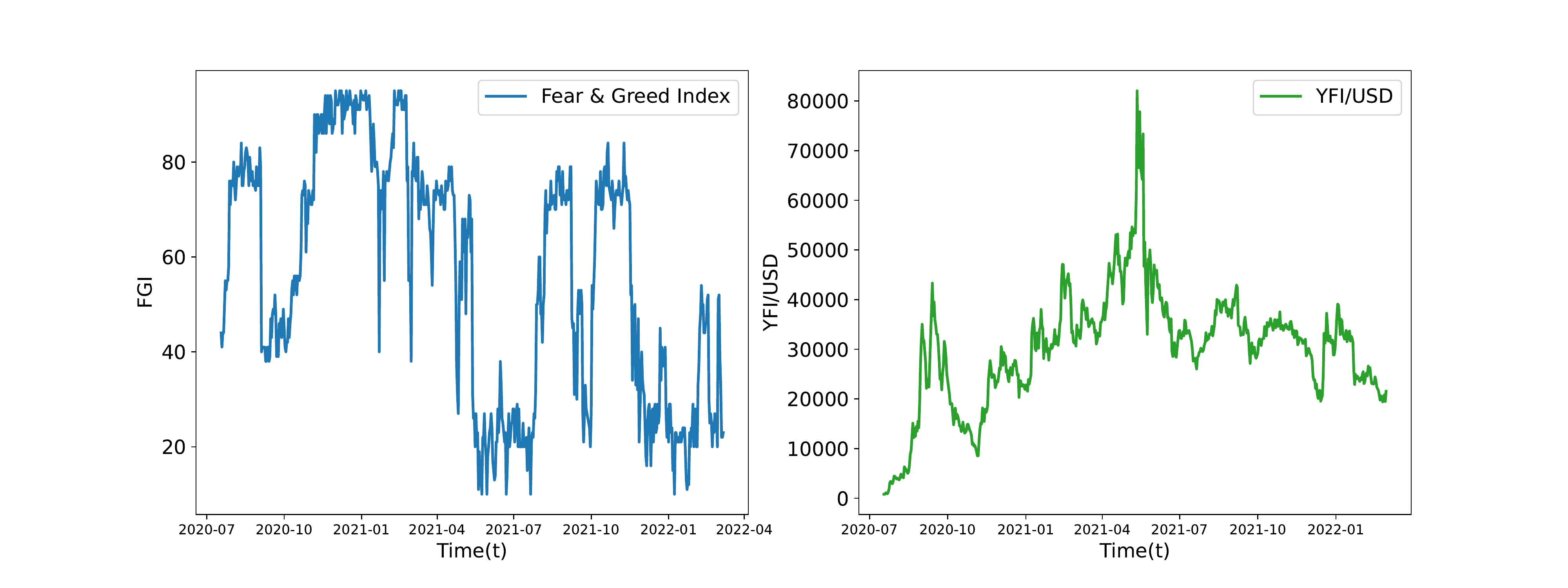}
    }
    \caption{Graphs for the Crypto FGI and YFI price.}
    \label{fig:price_FGI}
\end{figure}

%\subsection{Method}
\ac{DeFi} platforms, for the most part, are built on public-permissionless ledgers. Generally, these ledgers provide a rich source for the collection and analysis of quantitative data \citep{barbereau_decentralised_2022}. \citet{chen_blockchain_2020} observe that 80\% of \ac{DeFi} platforms, are in fact, built on the Ethereum ledger. Ethereum records a variety of details, not least on tokens, data about their creation, initial distribution to exchanges, and transaction history between addresses. The fair launch was originally created as part of Yearn Finance, hence it's practice informs our study inductively \citep{beese2019simulation}. Specifically, data on Yearn Finance is used to (1) define the base scenario $S_0$ ("Cronje"), (2) 'feed' our model based on reality, (3) calibrate the model, and (4) validate our model. For (2), we also extracted the price of YFI (from \href{https://www.coingecko.com/}{CoinGecko.com}) and the 
\href{https://alternative.me/crypto/fear-and-greed-index/}{Crypto Fear \& Greed Index (FGI)}. The graphs for these two additional data sources are presented in Figure \ref{fig:price_FGI}.

Yearn Finance is built on Ethereum and uses the ERC-20 token standard for its voting rights token YFI. YFI was launched with a fixed supply with no early allocation of tokens to insiders. Instead, the initial supply of 30,000 tokens in circulation was distributed via a liquidity providing scheme. Users could earn YFI by supplying liquidity into three distinct pools, allowing every user, regardless of their initial capital or other restrictions, to earn a portion of YFI’s supply proportionate to the contributed liquidity. This allocation was originally described as a \textit{fair launch} \citep{cronje_podcast_2020}. 

To generate data for our model, we used \href{https://dune.com/}{Dune} to extract the addresses that hold YFI from Ethereum's public ledger. Then, we organized the data such that we could determine how many tokens are owned daily by each address. Finally, we excluded a number of address 'types' from the dataset:
smart contracts (since they never utilized their voting rights, despite holding YFI \citep{barbereau_decentralised_2022}); 
addresses holding YFI valued less than \$1 (since these rarely vote or trade their tokens owing to Ethereum's gas fees being significantly higher than the token's value), and; 
addresses used to burn tokens (e.g., 0x000…0000) (since no one controls them and YFI is effectively taken out of circulation).
Table~\ref{tab:data_extraction} presents our final data set.

\begin{table}[ht!]
    \centering
    \caption{Overview of data extraction.}
    
    \begin{tabular}{lc}
    \toprule
        \textbf{Extracted Addresses}        &  96,227 \\
        \textbf{Addresses used in Analysis} &  86,752 \\
        \textbf{Extraction Period}          &  2020-07-17 - 2021-08-15 \\
    \bottomrule
    \end{tabular}
    \label{tab:data_extraction}
\end{table}

Following the finalization of our data set, we utilized Exploratory Data Analysis (EDA) to determine the model's initial conditions and variables. We chose September 1st, 2020 (i.e., 45 days after the project's launch) as the starting date since at that point the Yearn Finance fair launch took place; in other words, all tokens were allocated to users, but the market's influence on the token distribution was still limited. Using the Anderson-Darling test \citep{anderson_asymptotic_1952} and the Akaike information criterion \citep{akaike_new_1974}, we identified that the probability distribution of the initial YFI allocation follows a Lomax distribution ($\lambda=0.4$, $\alpha=0.5$). Relying on the same methods, we found that the daily number of new addresses that hold YFI increases following an asymmetric Laplace distribution $ALap(0.71, 58,76)$.

%% file: sections/04_1_model.tex
\section{The model}
\label{sec:model}

The proposed model for initial allocations builds on an agent-based artificial cryptocurrency market (c.f. \citep[]{cocco2017_abm_crypto}). Subsequently, we describe the model in terms of the agents, the market rules, and the trading behavior. Then, we describe the initial token allocations of the three fair launch scenarios. Finally, we introduce the metrics used to evaluate concentration over time.

\subsection{The agents}

For our model, we take time steps $t \in \mathbb{N}_{+}=\{1,2,3,...\}$ which correspond to a day and a new trading round. The first time step in our model is at $t = 45$. For each time step, we define agents $i \in I$ as the addresses that hold voting rights tokens (TKNs) at the beginning of each trading round. The number of agents at time step $t$ is given by $N_A(t) \in \mathbb{N}_{+}$. At the beginning of each trading round, a subset of $I$ is selected to trade TKNs (the selection mechanism as well as the trading strategy of agents is described subsequently) and new agents (endowed \textit{solely} with fiat currency) enter the market with the desire of placing buy orders to acquire TKNs. The new agents entering $(N_A(t+1) - N_A(t))$ follows $ALap(0.71, 58,76)$ for every $t > 45$. In other words, at each trading round the 95\% \ac{CI} of the number of new agents is $[114, 119]$. We run our model for 347 days $(t=392)$ and the 95\% \ac{CI} for the final number of agents ($N_A(392)$) is $[47113, 50890]$.

%Individual agents $i$ belong to the population of active traders $N_A(t)=\{1;2;3;4;...\}$ for each step $t\in\mathbb{N}_+=\{1;2;3;...\}$. In each trading round, a limited number of agents $N_A(t)$ trade, though continuously 'new' agents endowed with fiat may enter the market (traders who wish to acquire TKNs). Importantly, the actual trading begins post fair launch -- that is, after the initial allocation of the token supply $T_s$ to a group of 'historical' agents at $t=45$. From our data, we deduce that these new agents join at $t+1$ following an Asymmetric Laplace distribution such that $N_A(t+1>45) \sim ALap(0.71, 58,76)$ returning values between [114, 119] with (95\% CI). From the data we also defined the simulation length: the simulation is completed at $t=392$ (or after 347 days). Correspondingly, the final number of agents is $62808 \leq N_a(392) \leq 65189$ (95\% CI).\red{recompute the CI for t = 347}

Agents are endowed with fiat holdings $f_i(t)$ and TKN holdings $y_i(t)$. Based on \citet{dragulescu_exponential_2001} and \citet{brzezinski_wealth_2014}, the amount of fiat held by both, individual agents at $t=45$ and those agents entering the market at each trading round, is drawn from a Pareto distribution with $\alpha=2.1$ and $min(f_i(t))=\$400k$ for the richest 10\% of our agents and from an $Exp(\frac{1}{40000})$ for the remaining bottom 90\%. The amount of TKNs held by agents at $t=45$ depends on the chosen fair launch scenario $S_i$  with $i\in\{0,1,2\}$. 

Independently of fiat or TKN holdings, each agent $i$ is assigned to one of two populations, Diamond Hands (DH) and Random Traders (RT), representative of respective trading strategies. DHs are risk averse traders, who pragmatically invest in the market and are more likely to not incur in trades. RTs, then, are agents who enter to market for a variety of reasons (e.g., portfolio diversification, gambling, etc.). Following \citet{cocco2017_abm_crypto}, the agent populations is divided into 30\% DHs and 70\% RTs.

%%%%%%%%%%%%%%%%%%%%%%%%%%%%%%%%%%%%%%%%%%%%%%%%%%%%%%%%%%%%%%%%%%%%%%%%%%%%%%%%%%%%
\subsection{The market rules}

%While our study was informed by the \ac{ABM} of \citet{cocco2017_abm_crypto}, instead of the order book mechanism for transaction clearing used in their work, we opted for a mechanism comparable to the principles of a clearing house such that buy and sell orders are matched periodically \citep{mendelson1982market}. The clearing mechanism we utilize though is \textit{not} a clearing house, because it does not account for price formation, nor does it include adjustments of price after every transaction over time but instead the TKN price is updated daily and remains stable throughout the day. Specifically, the clearing mechanism allows to address RQ1 and determine which market conditions affect the concentration of voting rights tokens. As a proxy to simulate the market conditions, we use the Crypto Fear \& Greed Index composed of volatility, market momentum/volume, social media, dominance, and trends \citep{FGI_2022}. 

The TKN market is given by a mechanism \textit{comparable} to a clearing house; whereby, buy and sell orders are accumulated over time and cleared ('matched') periodically \citep{mendelson1982market}. The purpose of the model and developed market is not exploring how price is formed; instead, it is to simulate how tokens circulate (and concentrate) based on clear conditions. The mechanism we utilize is \textit{not} a formal clearing house as it does not account for price formation, nor does it include adjustments of price after every transaction over time. Instead, at each time step the TKN price $T_p(t)$ is updated based on YFI's historical price data. Agents can autonomously decide whether they are willing to trade. Agents do not, however, have information about the orders other agents are placing. 
%\reviewer{Extend your paper with solid arguments for and against the use of a clearing house model in your simulations }
The scope of this work and the developed \ac{ABM} is on token concentration, and not the way price is formed. Clearing houses offer an easy to deploy mechanisms to match orders between agents with limited computation overhead and a realistic movement of the tokens.

%We evaluate the centrality of the model after the token supply $T_s$ is distributed among $N_A$ and agents are able to trade. We take all the buy/sell orders created by the agents to have a fixed price at every time step. The TKN price $T_p(t)$ is given by historical data from the Yearn Finance price. After the initial token allocation, $T_p(45)=\$35059$. 

At $t\geq45$, the total number of tokens in circulation is given by the constant $T_s=36666$. For the trade of TKNs, we model a two-sided market with a number of buyers, each willing to buy TKNs, and several sellers, each willing to sell TKNs. Additionally, at every time step the buy/sell orders created by the agents are matched in a \textit{first in first out} method, and at the end, the unmatched orders are canceled. 
%The TKNs are homogeneous, all agents act independently, and each has their own probability $P_i(T)$ to engage in a trade.

%For the trade of TKNs, we model a two-sided market with a number of buyers, each willing to buy TKNs, and several sellers, each willing to sell TKNs. The TKNs are homogeneous, all agents act independently, and each have their own probability $P_i(T)$ to engage in a trade. 
%In doing so, we developed a market whereby orders are matched following principles of a simplified clearing house \citep{mendelson1982market}\opnote{do we need this sentence? because we also mention it at the beginning of the section}. 

%%%%%%%%%%%%%%%%%%%%%%%%%%%%%%%%%%%%%%%%%%%%%%%%
\subsection{Trading behavior}

%\subsubsection{Trade or no trade?}%first: trade or no trade?
Depending on the population agents belong to, they exert a choice -- to trade ($T$) or not to trade -- at every time step $t$. This decision is given by the probability $P_i(T)$. For RTs, who randomly wish to trade following a uniform distribution, $P_i(T)=0.5$. For DHs, $P_i(T)$ is dependent on two independent variables.
First, the Fear \& Greed Index ($FGI(t)$) which fluctuates between a value of 0 (``Extreme Fear'') and 100 (``Extreme Greed'') \citep{FGI_2022}. In our case, we consider the values of the index as ``Extreme'' ($FGI_e$) when  $FGI(t)>th_{h}$ or $FGI(t)<th_{l}$, and ``Normal'' ($FGI_n$) when $th_{h}>FGI(t)>th_{l}$ with $th_{h}$ and  $th_{l}$ thresholds for the extreme values of Fear \& Greed Index.
Second, the agent's wealth ($W$), given by an agent's individual holding denominated in fiat $f_i(t)$. An agent's wealth at time $t$ is considered ``High'' ($W_h$) when $f_i(t)$ is above the 90th percentile of the wealth distribution and ``Low'' otherwise. 
Therefore, the probability of a DH agent to trade is given by: 

\begin{equation}
\begin{split}
    P_i(T) &= P(T\|FGI_e, W_h)P(FGI_e)P(W_h) + P(T\|FGI_e, W_l)P(FGI_e)P(W_l)\\ 
            & +  P(T\|FGI_n, W_h)P(FGI_n)P(W_h)  + P(T\|FGI_n, W_l)P(FGI_n)P(W_l)
\end{split}
\end{equation}

%where $FGI$ represents the mentioned Fear \& Greed Index and $W$ represents wealth and $T$ represents the likelihood of a agent to incur in a trade. \tbnote{we dont need this. its exlpained above}

%\subsubsection{Buy or sell?}%second: buy or sell?
If an agent is willing to trade, the subsequent decision to execute a buy or sell order depends on the population they belong to. For RT, the buy and sell orders follow a Bernoulli distribution with $p = 0.5$. Initially, the same holds for DH but in their case the probability is calibrated at a later stage based on the data from Yearn Finance (c.f. Section \ref{sec:cali}). In the trading behavior, we do not consider protocols that allow to stake/sell voting rights token entitlements (e.g., Bribe Protocol) as these add yet another degree of complexity.

%\subsubsection{How much to buy or sell?}%third: how much? 
At each time step the amount of fiat an agent spends on buying tokens follows a $\mathcal{N}(\mu=\frac{f_i(t)}{2},\,\sigma=\frac{\mu}{3})$ and the number of TKNs an agent sells follows a $\mathcal{N}(\mu=\frac{y_i(t)}{2},\,\sigma=\frac{\mu}{3})$. 
%\reviewer{The behaviour of small YFI holders. In practice, users can stake their governance tokens to 'voting buying' protocols, e.g., Bribe Protocol (https://www.bribe.xyz/). Small holders may not even trade their governance tokens but make profits by using such services. So, I feel the model is not very close to what is happening.}DONE - I wrote a disclaimer about it...

%%%%%%%%%%%%%%%%%%%%%%%%%%%%%%%%%%%%%%%%%%%%%%%%
\subsection{The fair launch scenarios}

Our simulation is set up around three distinct scenarios representative of initial token allocations understood as 'fair'. Their design is informed on the basis of the epistemic dichotomy described in \citet{beese2019simulation}. The base scenario $S_0$ is created following an inductive approach (its design is informed by data extracted from Yearn Finance) and the distribution of $T_s$ is modeled to follow a Lomax distribution with $\lambda = 0.4$ and $\alpha = 0.5$. The artificially created fair launch scenarios $S_1$ and $S_2$ are designed following a deductive approach on the basis of theory. 

The first alternative scenario $S_1$ (``Bentham'') considers 'fairness' in egalitarian terms: equity is achieved in terms of uniformity such that the total supply of tokens is divided equally among the participants. Formerly, it considers Jeremy Bentham's dictum that ``everybody to count for one, nobody for more than one'' \citep{mill_1864}, without consideration of individual interests or material situation. For $S_1$, $T_s$ is uniformly distributed such that each agent $i$ at $t=45$ holds $y_i(45)=\frac{T_s}{N_A(45)}$. 

The second alternative scenario $S_2$ (``Rawls'') considers randomness, and more specifically, the principle of a lottery as 'fair': equity is achieved in terms of a token allocation -- at random --, and in our case, following a Normal distribution. It re-hashes the idea that the outcome of each individual's position, like the outcomes of ordinary lotteries, is a matter of good or bad ``luck'' \citep[p. 74-5]{rawls1971theories}. Randomness and chance are central to the theory of Darwinian evolution \citep{wagner2012role}. For $S_2$, $T_s$ is distributed among agents following a truncated Normal Distribution ($\mu=0.103,\sigma=0.192$) defined on $[0, \infty]$.

In sum, we thus investigate two additional scenarios aside (Table \ref{tab:scenarios}). While keeping the market conditions and parameters fixed, changing the initial allocation of tokens provides further insight into the concentration of wealth. 
%\reviewer{Please mention in the paper more explicit which scenarios and trading behaviours your simulation models cover and which not. Reviewer 2 pointed out some examples of how governance tokens can be used for other activities. I see the merit of your approach and final results, though it clearly has certain drawbacks. Therefore, please be more transparent and specify the scope of your simulation models in order to curb readers’ expectations. }DONE (see last paragraph of the calibration)

\begin{table}[thb]
    \centering
    \caption{Initial token allocation scenarios.}
    \begin{tabular}{llll}
    \toprule
        &\textbf{Scenario}    & \textbf{Allocation}  & \textbf{Perspective} \\ 
    \midrule
        $S_0$ & ``Cronje'' (Yearn Finance) & Everyone gets the same opportunity & Social liberalism \\  
    %\hline
        $S_1$ & ``Bentham''                & Everyone gets the same             & Egalitarianism    \\ 
    %\hline
        $S_2$ & ``Rawls''                  & Everyone gets a random amount      & Darwinism         \\ 
    \bottomrule
\end{tabular}
\label{tab:scenarios}
\end{table}

%% file: sections/04_2_metrics.tex
\subsection{Metrics}
\label{sec:metrics}

Given the aim of analyzing the distribution of voting rights tokens post fair launch, select metrics are computed at every time step. These metrics are the Gini Coefficient \citep{1912vamu.book.....G} and the Shannon Entropy \citep{shannon_mathematical_1948}. This choice was made based on an evaluation of related works seeking to quantify and measure the distribution of tokens in a system; notably, as discussed in \citet{gervais_is_2014}, \citet{gochhayat_measuring_2020}, and \citet{barbereau2022defi:HICSS}. 

The Gini Coefficient is typically used to assess the distribution of wealth in a given country. It was, however, also applied to study wealth distribution in Bitcoin and Ethereum \citep{gochhayat_measuring_2020} as well as \ac{DeFi} platforms and voting rights tokens distributions \citep{barbereau2022defi:HICSS}. For our model, the Gini $G$ indicates the concentration of wealth, and in particular the concentration of TKNs amid agents. The Gini is given by: 
\begin{equation} \label{eq:Gini}
G= \dfrac{\sum\limits_{i=1}^{N_A}\sum\limits_{j=1}^{N_A}|p_i-p_j|}{2N_A\cdot\sum\limits_{j=1}^{N_A}p_j}
\end{equation}
where $p_i$ corresponds to the share of TKNs held by agent $i$ and $N_A$ the total number of agents. It is maximized through the Dirac distribution $\delta_{i_0}$, i.e., $p_{i_0}=1$ for some $i_0\in\{1,\ldots,N_A\}$ and $p_i=0$ for all $i\neq i_0$, and minimized through the uniform distribution, i.e., $p_i=\frac{1}{N_A}$ for all $i$.

The Shannon Entropy was developed to assess the information loss in telecommunication networks \citep{shannon_mathematical_1948}. The Normalized Shannon Entropy (NSE), then, takes values between 0 and 1, and determines the unpredictability of a distribution. We assume that a system where the voting tokens are distributed can exhibit high unpredictability (1), given that more agents influence the outcomes. The NSE is given by:
\begin{equation} \label{eq:NSE}
\text{NSE}=-\sum\limits_{i=1}^{N_A}\dfrac{p_i\log(p_i)}{\log N_A}
\end{equation}
where $0\log(0)\equiv0$ by convention since 
$\lim\limits_{p \to 0} p\log(p) = 0$.
It is~0 for $\delta_{i_0}$ and 1 for the uniform distribution (i.e., the extremes are interchanged compared to the Gini coefficient).
To ease graphical observation, we opted to consider 1-NSE instead of NSE such that, as in Gini, higher values correspond to higher degrees of centrality.

%% file: sections/04_3_calibration.tex
\section{Implementation and Calibration}
\label{sec:cali}

%\reviewer{Although I liked the background and motivation, it was a bit lengthy. I think a shorter and more concise background could improve the reading experience. The additional space would be well-used in the description of the data-set of the calibration method.}

The model was implemented in Python using the MESA framework \citep{python-mesa-2020}. The simulation took about 4.5 hours in a 10 Cores M1-Pro CPU with 32GB of RAM. Due to this computational burden. The calibration of the model was performed in a High Performance Computing (HPC) facility. %~\Hide{\citep{HPC}}.
The hardware provided, depending on the allocation of the HPC, were a Dual Intel Xeon Broadwell or Skylake with 128GB of RAM.

%In order to generate a trustworthy model, we take the scenario $S_0$ and calibrate patterns to mimic the extracted data of Yearn Finance. We do so for the buy/sell ratio, the agent distribution, and probabilities. 

%\subsection{Optimization of parameters}

For the calibration, we followed the recommendations of \citet[p. 4]{richiardi2006common} whereby a “full exploration” of the parameters is required. To do so, we implemented a grid search (GS) to find a set of optimal values of parameters. GS performs an exhaustive search over all the possible combinations of parameters until finding the optimal one.
The goodness of the fit and the stopping condition of GS are computed using Root Mean Squared Error (RMSE) and Mean Absolute Percentage Error (MAPE) between the actual (extracted from the dataset) and calibrated model. The respective equations are given by:
\begin{equation} \label{eq:optimization}
\text{RMSE} = \sqrt{\left(\frac{1}{n}\right)\sum_{i=1}^{n}(x_{i} - \hat{x_i})^{2}} \quad \text{and} \quad \text{MAPE} = \frac{100}{n}\sum_{t=1}^n \left | \frac{x_{i}-\hat{x_i}}{x_{i}} \right |
\end{equation}
where $x_i$ is the actual observation and $\hat{x_i}$ is the simulated value.

%\subsection{Parameter optimization}
The optimization was executed over all eligible parameters. The FGI threshold takes values from 0 to 100. For all other parameters, we considered values between 0 and 1. The optimal parameter values are displayed in Table \ref{tab:parameter_optimization}.

\begin{table}[h!]
\centering
\caption{Optimal parameter values.  \label{tab:parameter_optimization}}
\resizebox{\columnwidth}{!}{%

\begin{tabular}{ccc}

\toprule
\textbf{Parameter}  &\textbf{Description}                               & \textbf{Optimal Value} \\ \midrule
$P_{DH}(Buy)$        & Buy probability of DH                                    &   0.7                  \\ 
$DH/N_A(t)$           & Population share of DH                       &   0.3                  \\ 
$th_{h}$        & FGI threshold high                                    &   80                   \\ 
$th_{l}$        & FGI threshold low                                     &   20                   \\ 
$P(T\|FGI_e)$   & Trading probability under extreme market conditions   &   0.7                   \\ 
$P(T\|W_h)$     & Trading probability under high wealth                 &   0.7                  \\ 
$P(T\|FGI_n)$   & Trading probability under normal market conditions    &   0.8                   \\ 
$P(T\|W_l)$     & Trading probability under low wealth                  &   0.9                   \\ \bottomrule

\end{tabular}
}
\end{table}

The optimal values of the DH/RT population ratio were close to the ratio used by \citet{cocco2017_abm_crypto}. Therefore, we fixed it at 30\% DHs and 70\% RTs. Similarly, the buy probability was optimized to be 70\% for DH. 
%For the RTs, the probabilities are represented by a constant such that $P(FGI_e) = P(W_h) = P(FGI_n) = P(W_l) = 0.5$. 
For the DHs, we found that high trading probabilities indeed lead to lower error rates (Table \ref{tab:error_rates}). This, as demonstrated in \citet{rocsu2021evolution}, represents an expected behavior as more trading is linked with higher wealth concentration. Diametrically opposed to the high trading probability parameter set, is an artificially created parameter set, with relatively low trading probabilities. The error rates for this parameter set are relatively worse than the optimal set of high trading probabilities. We also artificially generated and investigated a compromise between the two sets without extreme trading probabilities.

\begin{table}[ht!]
\centering
\caption{Diamond Hand trading probabilities parameter sets with their correspondent error rates  \label{tab:error_rates}}
\resizebox{\columnwidth}{!}{%

\begin{tabular}{cccc}
\toprule
\textbf{Parameters}  & \textbf{High probability} & \textbf{Medium probability} & \textbf{Low probability} \\ \midrule
$P(T\|FGI_e)$       & 0.7               & 0.3        & 0.1                   \\ 
$P(T\|W_h)$         & 0.7               & 0.4        & 0.1                   \\ 
$P(T\|FGI_n)$       & 0.8               & 0.3        & 0.2                   \\ 
$P(T\|W_l)$         & 0.9               & 0.5        & 0.2                   \\  
$P_i(T)$         & 0.77            & 0.38        & 0.15                 \\ \midrule
$MAPE$   & 0.1859               & 0.224        & 0.255                   \\  
$RMSE$   & 0.007               & 0.009        & 0.012                  \\  

\bottomrule

\end{tabular}
}
\end{table}

%Assuming half of the time the market is under 'extreme conditions ($FGI_e$) and half of the time in 'normal' conditions ($FGI_n$), the probability of trading assuming 50\% of the agents are high wealth and 50\% low wealth. The resulting trading probability is 60\% for High trading probability and 2.25\% for low trading probability.

%To evaluate the robustness of the agent distribution amid the two populations - DH and RT - we considered alternative settings to the initialized 30/70 distribution of \citet{cocco2017_abm_crypto} model. We evaluated the impact of alternative ratios, namely 50/50 and 10/90, on the metrics. After the optimization, we do not denote a significant impact on the metrics. Therefore, we subsequently continue the simulations with a fixed allocation of 30/70 (like in \citet{cocco2017_abm_crypto}).

In sum, we investigate three scenarios (Table \ref{tab:scenarios}) under three trading probabilities (Table \ref{tab:error_rates}). The results might vary due to the stochastic nature of \ac{ABM}. In anticipation of this variance and to ensure the robustness of our results, we applied a Monte-Carlo method by repeating the experiment of the three simulation sets within the HPC; resulting in more than 1000 simulations (or, approximately 300 per set of trading probabilities).
For all simulations, the agents can place buy or sell orders depending on the probability defined in $P_{DH}=0.7$ for DH (optimized value) and $P_{RT}=0.5$ for RT (constant) respectively.

%% file: sections/05_results.tex
\section{Simulation results}
\label{sec:results}

%Previously, we defined a suitable set of parameters that minimize the error between the extracted and simulated metrics. In this section, we explore the effects of different parameters on the behavior of our model in terms of three simulation sets representative of high, medium, and low trading probability respectively (Table \ref{tab:error_rates}). The trading probability defines the likelihood that an agent places a trading order. For all simulations, the agents can place buy or sell orders depending on the probability defined in $P_{DH}=0.7$ for DH (optimized value) and $P_{RT}=0.5$ for RT (constant) respectively. In each simulation with respective probabilities, we consider the three scenarios $S_0$, $S_1$, and $S_2$. \tbnote{Joa: i move parts of this paragraph up, but cut most of it. no need to say 5 times what we do.}

%\citep{HPC}. 

%%%%%%%%%%%%%%%%%%%%%%%%%%%%%%%%%%%%%%%%%%%%%%%%%%%%%%%%%%%%%%%%%%
\subsection{Effects of trading probability on the three scenarios}

%In this section, we will depict the results of the simulations under different trading probabilities and different initial allocations namely scenarios $S_0$, $S_1$, $S_2$. 

\begin{figure}[H]
    \centering
    \makebox[0pt]{
    \includegraphics[width=0.9\columnwidth]{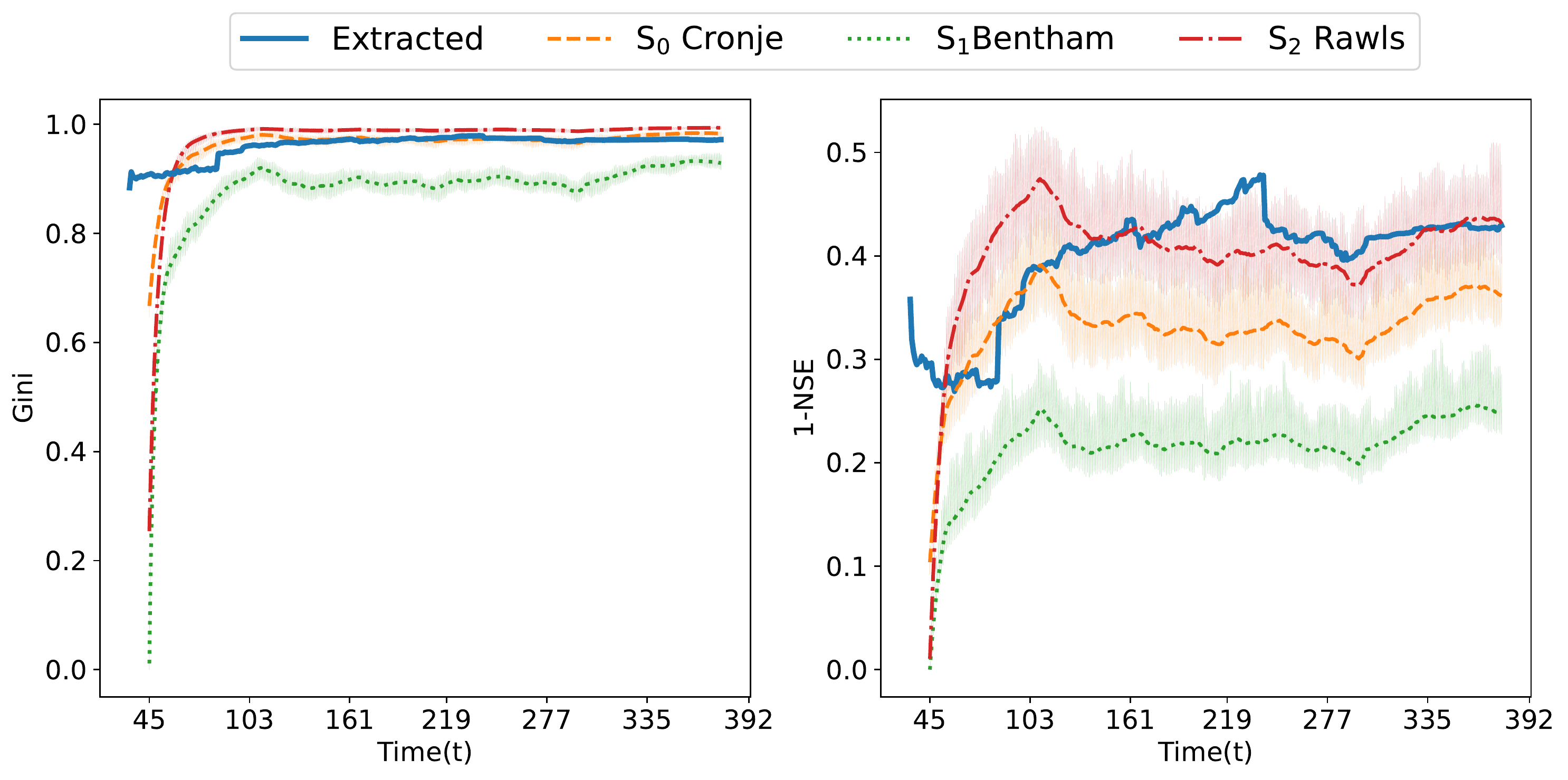}
    }
            \caption{Simulations for the parameter set representative of high trading probabilities.}

    \label{fig:high_prob_scenarios}
\end{figure}

\begin{figure}[H]
    \centering
    \makebox[0pt]{
    \includegraphics[width=0.9\columnwidth]{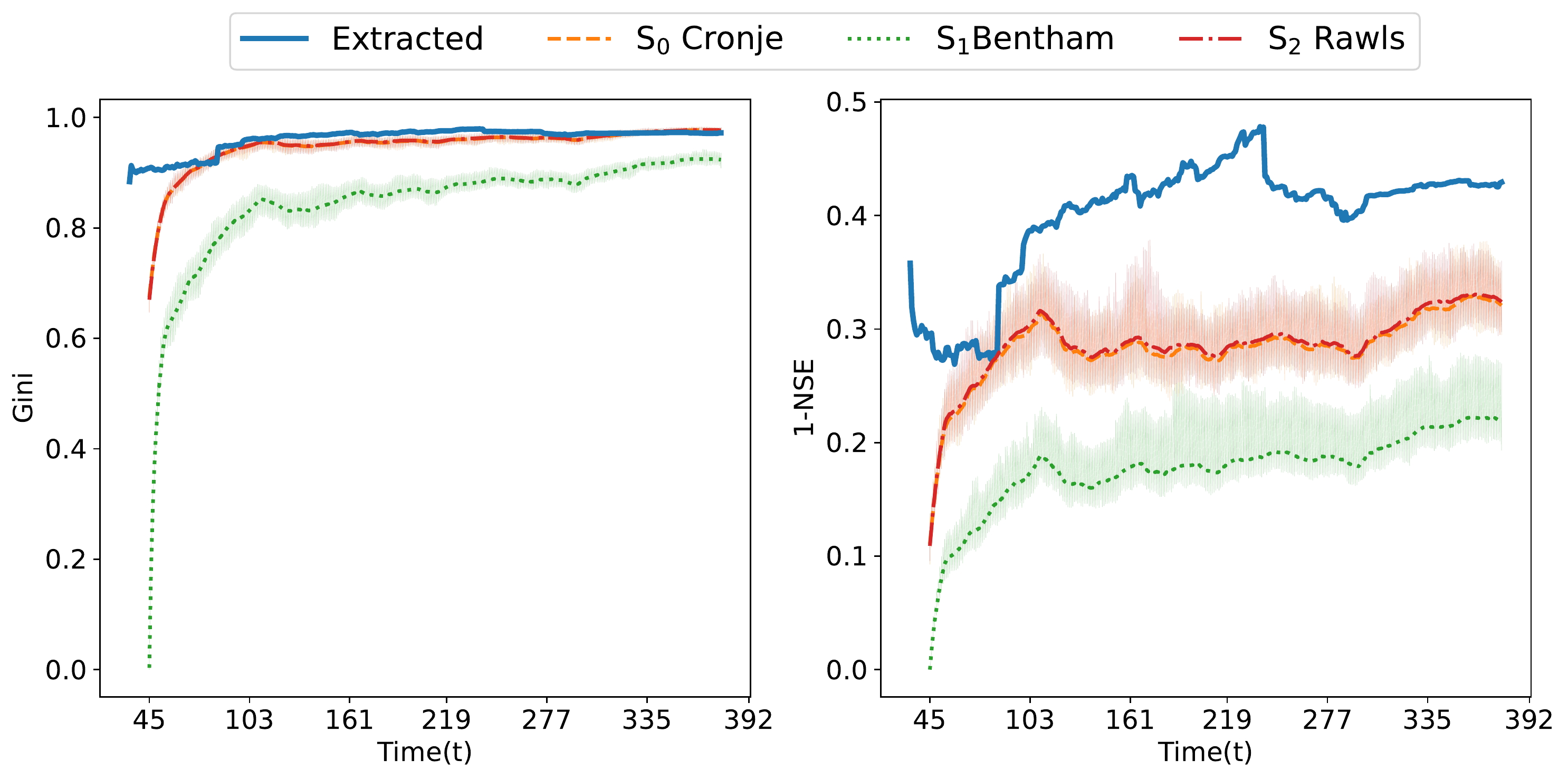}
    }
    \caption{Simulations for the parameter set representative of low trading probabilities.}
    \label{fig:low_prob_scenarios}
\end{figure}

\begin{figure}[H]
    \centering
    \makebox[0pt]{
    \includegraphics[width=0.9\columnwidth]{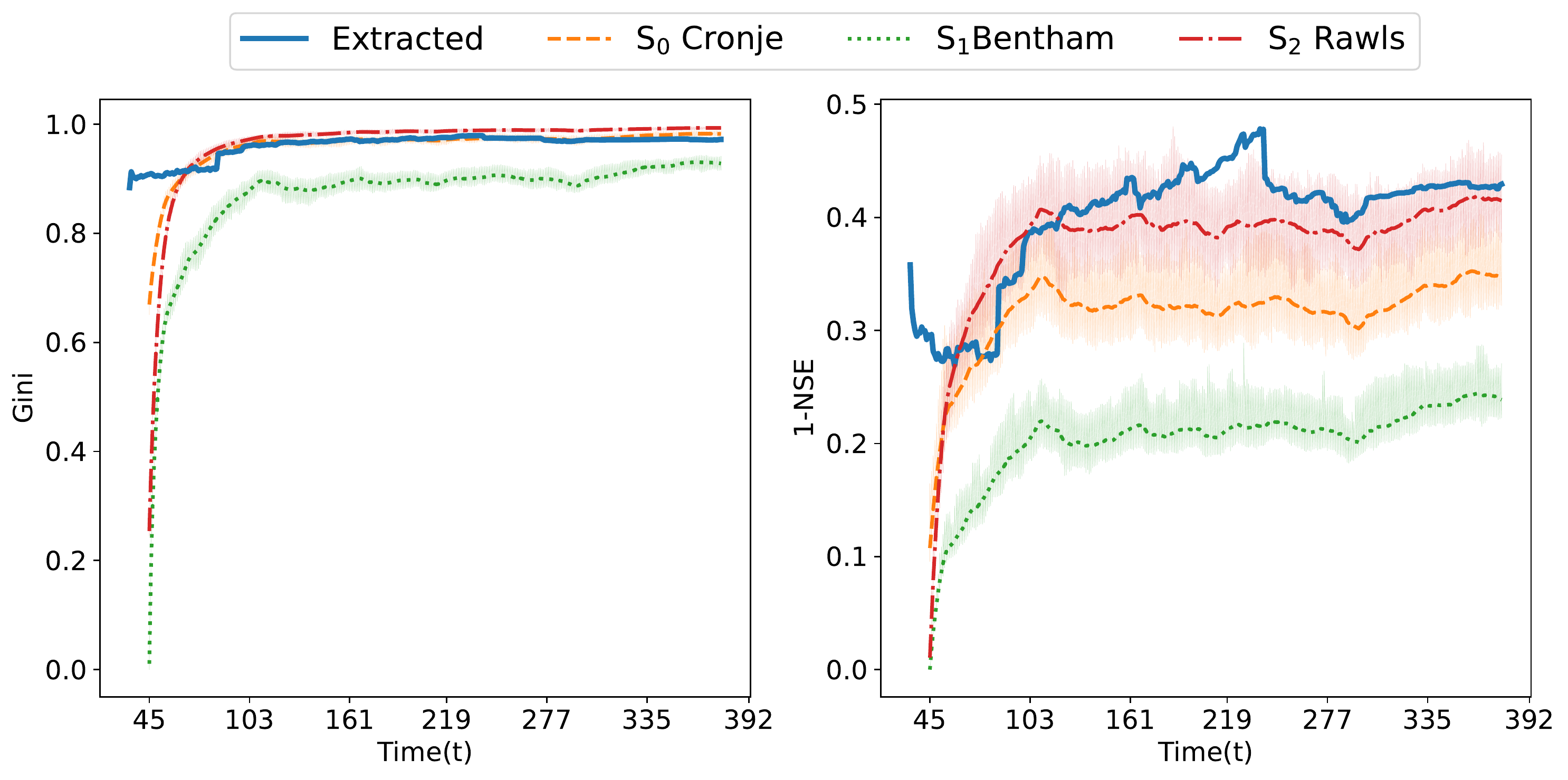}
    }
    \caption{Simulations for the parameter set representative of medium trading probabilities.}
    \label{fig:medium_prob_scenarios}
\end{figure}

The first simulation considers the model's behavior under high trading probability (Figure \ref{fig:high_prob_scenarios}). High trading probability refers to a relatively high likelihood for DH agents to place an order.  
The second simulation is the artificially created edge case with low trading probabilities (Figure \ref{fig:low_prob_scenarios}). It is diametrically opposed to the former simulation and explores the behavior of DH agents when the market dictates a relatively low likelihood to place a trade. (The graphs for $S_0$ and $S_2$ are visually coinciding.)
The third simulation set was created artificially as middle ground between the high and low trading probabilities (Figure \ref{fig:medium_prob_scenarios}).

Across all trading probabilities, for $S_0$ and $S_2$, the Gini values of the simulated data overlap considerably with the extracted YFI data for $t\ge100$. Until the end of the simulation time frame, the Gini values of the extracted data diverge by at most by 1.1\% on $S_0$ and 1.7\% on $S_2$ across all trading probability scenarios. The close fit between the extracted data from Yearn Finance and the three simulations, and in particular for $S_0$, is expected given the performed optimizations. The exception to the convergence is $S_1$, whose graph is below the extracted data in all three simulations. It appears that an egalitarian initial token allocation would then lead to relatively less concentration in the time frame of the simulation.

We expect variations in the initial values of Gini and NSE because, even though our model starts with the same number of agents at each simulation, the initial token allocation is not fixed and as discussed above, the token distribution in $S_0$ follows a Lomax distribution.
%1-NSE is a measure of randomness and, in the context of tokens, a measure of decentrality. Higher values of 1-NSE display high similarity in the total tokens held ($y_i$) by the agents. After the initial sharp rise due to agents' will to acquire new tokens, the metric grows steadily as agents without token holdings enter in the simulation. 
%On the difference scenarios we can see how the initial token distribution set a ceiling up to which the 1-NSE can rise and maintain a parallel move between then and the extracted 1-NSE. 
%Within the different scenarios, we observe how the distribution imposes a ceiling on how high the 1-NSE might grow. 
Regardless of the scenario and trading probability, after the 1-NSE stabilizes, all distributions move in a lateral and parallel direction with regards to the extracted 1-NSE values from YFI.

Although  our simulation results seem to coincide with the Yearn Finance data for Gini we observe high variations of the NSE. 
%At around $t=120$, for all simulations, we observe a divergence between the extracted and simulated NSE. 
This divergence may be interpreted in two ways.
At inception of Yearn Finance some contracts held a large amount of YFI and distributed them shortly after. Given the scope of our model we did not consider such behavior.  We argue that the smart contracts that emitted YFI rapidly result in sharper rises in the metrics' values. This is consistent with our findings which indicate that when there is a large amount of YFI accessible for trading in a short period of time, the metrics rise.
The second interpretation pertains to the token supply ($T_s$). Formerly, the supply of YFI was ``schedule-based'' \citep{oliveira2018token}: it began with a supply of 30000 FYI allocated following the fair launch, and subsequently an additional 6666 YFI tokens were distributed the same way. For simplicity, we start with 36666 supply dispersed to the starting holders in our simulations. Again, the distribution of 6666 YFI in a short period of time theoretically results in higher concentration than our model, which in contrast, distributes YFI more slowly over time. %This trend can be observed across all scenarios: the graph begin at a constant and smoothly rise.
The schedule-based supply of YFI can be observed in the 'bumps' around t=100. While the change is more subtle in Gini, within 1-NSE the change is more clearly observable. This is due to the comparatively higher sensitivity of the latter metric with regards to minor fluctuations \citep{barbereau_decentralised_2022} which can also be observed in the Figures above, where the standard deviation of NSE is substantially higher than that of Gini.

%With regards to the Gini under low trading probabilities, like under the high trading probabilities, we observe a similar convergence of $S_0$ and $S_2$ towards that of Yearn Finance. The exception, again, being $S_1$ where G stabilizes around 0.9. For the NSE, \tbnote{joaquin/orestis: can you continue the interpretation on NSE?}. 

%%%%%%%%%%%%%%%%%%%%%%%%%%%%%%%%%%%%%%%%%%%%%%%%%%%%%%%%%%%%%%%%%%
\subsection{Actual concentration of wealth amid \textit{whales}}

Following the three simulation sets focusing on the trading probabilities, we performed a more granular analysis of the actual concentration of TKNs amid the population of agents ($N_A(392)$). Particularly, we sought to investigate the share of agents that hold 90\% of all tokens in circulation. These agents are so called \textit{whales}, `` `wealthy', above-average token-holders'' \citep[p. 20]{barbereau_decentralised_2022}. In consideration of the amount of available data following the Monte-Carlo simulations, here we present a more feasible analysis on the basis of the results from the first simulation round. The results are presented in Table \ref{tab:whales}. 

\begin{table}[ht!]
\caption{Share of agents that control 90\% of TKNs in circulation at t=392.\label{tab:whales}}

\centering
\resizebox{\columnwidth}{!}{%

\begin{tabular}{ccccccc} 
\toprule
\multicolumn{1}{c}{}                  & \multicolumn{2}{c}{\textbf{High probability}} & \multicolumn{2}{c}{\textbf{Medium probability}} & \multicolumn{2}{c}{\textbf{Low probability}}   \\ 
\midrule
\multicolumn{1}{c}{\textbf{Scenario}} & \multicolumn{1}{c}{Percentage} & Actual       & \multicolumn{1}{c}{Percentage} & Actual         & \multicolumn{1}{c}{Percentage} & Actual        \\ 
\cmidrule(lr){2-3}\cmidrule(lr){4-5}\cmidrule(lr){6-7}

\textbf{S0} Cronje                      & 2,59\%                          & 1137 / 43830 & 2,63\%                          & 1188 / 45092   & 3,63\%                          & 1499 / 41248  \\ 

\textbf{S1} Bentham                     & 10,80\%                         & 4777 / 44214 & 10,73\%                         & 4847 / 44950   & 11,38\%                         & 4999 / 43895  \\ 

\textbf{S2} Rawls                       & 0,76\%                          & 376 / 49397  & 1,29\%                          & 572 / 44300     & 2,83\%                          & 1250 / 44056  \\
\bottomrule
\end{tabular}
}
\end{table}

Unsurprisingly, in consideration of the values metrics took in the previous analysis, we observe a concentration of TKNs in the hands of the few. These few individuals are de facto in control as they may exert significant political pressure. In relative terms, as reflected in the metrics, the egalitarian allocation $S_1$ shows that the actual number of whales is higher. Regardless, our results are aligned with the timocratic description of \citet{barbereau2022defi:HICSS}.

%%%%%%%%%%%%%%%%%%%%%%%%%%%%%%%%%%%%%%%%%%%%%%%%%%%%%%%%%%%%%%%%%%
\subsection{Extending the simulation of S\_1 ``Bentha''}
\begin{figure}[ht!]
    \centering
    \makebox[0pt]{
    \includegraphics[width=0.9\columnwidth]{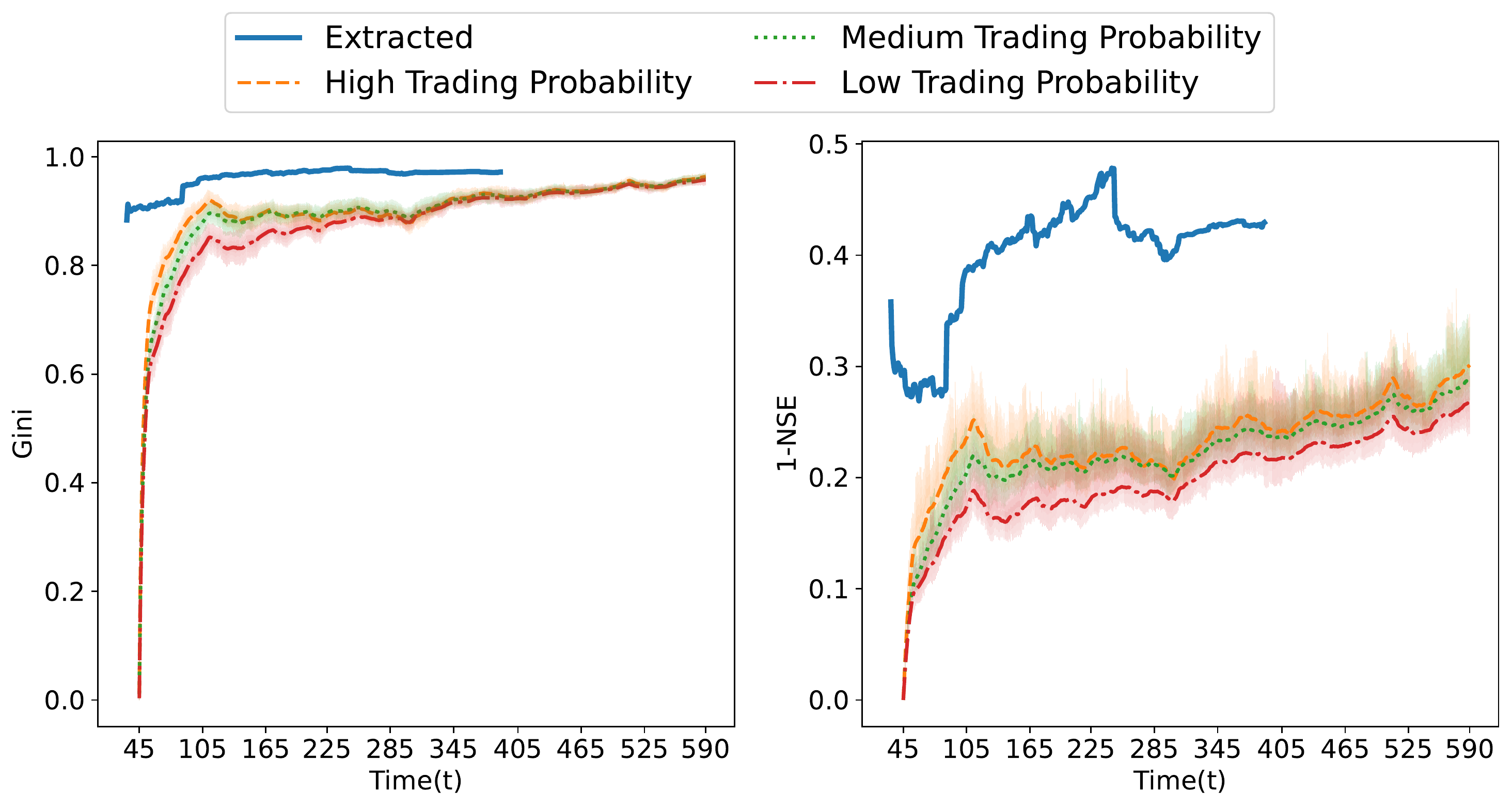}
    }
    \caption{Simulation for the Bentham scenario under the three trading probabilities.}
    \label{fig:bentham_scenarios}
\end{figure}

After running the first set of simulations, we opted to run a separate simulation to explore whether $S_1$ indeed demonstrates more or less concentration over time. To do so, we extended the simulation rounds from t=392 (August 15th 2021, the last data point extracted from Yearn Finance) to t=545 (March 1st 2022, the last point of the simulations). This represents an extension of 44.39\%. 
The results of the simulation are in Figure \ref{fig:bentham_scenarios}. 
%\reviewer{Figure 5 why is the yearn data cut off? "this extension of 44.39\% terminates the analysis at March 1st 2022 (the final data point extracted from Yearn Finance" suggests to me that both lines should be extend to the end of the observation period.} DONE

From our previous simulations on the effects of different trading probabilities, we observe how $S_1$ Bentham's initial token allocation positively affects both metrics. It is to be expected that an equal distribution of tokens at origination will reduce concentration, at least early on. In this simulation we observe similar phenomena to what was demonstrated in \citet{rocsu2021evolution}: even though the delayed effect that an egalitarian initial token allocation like the one simulated might generate, concentration is, at least judging from our simulations, inevitable in the long run. To corroborate that, we fitted linear regressions (LR) on Gini from the three trading scenarios. In the worst case scenario, the slope of the LR is $4*10^{-4}$.

%\reviewer{This work misses a thorough discussion on how to validate the simulation result.}DONE

%% file: sections/05_verification.tex
\section{Validation and verification of the model}
\label{sec:validation}

Validation is an essential part of \ac{ABM} \citep{davis2007developing}. There are numerous techniques for validation, all of which are used to establish credibility in the simulations \citep{beese2019simulation}. To validate our model we opt to use the \textit{event validity} and \textit{parameter variability (sensibility analysis)} techniques.

\begin{figure}[ht!]
    \centering
    \makebox[0pt]{
    \includegraphics[width=0.9\columnwidth]{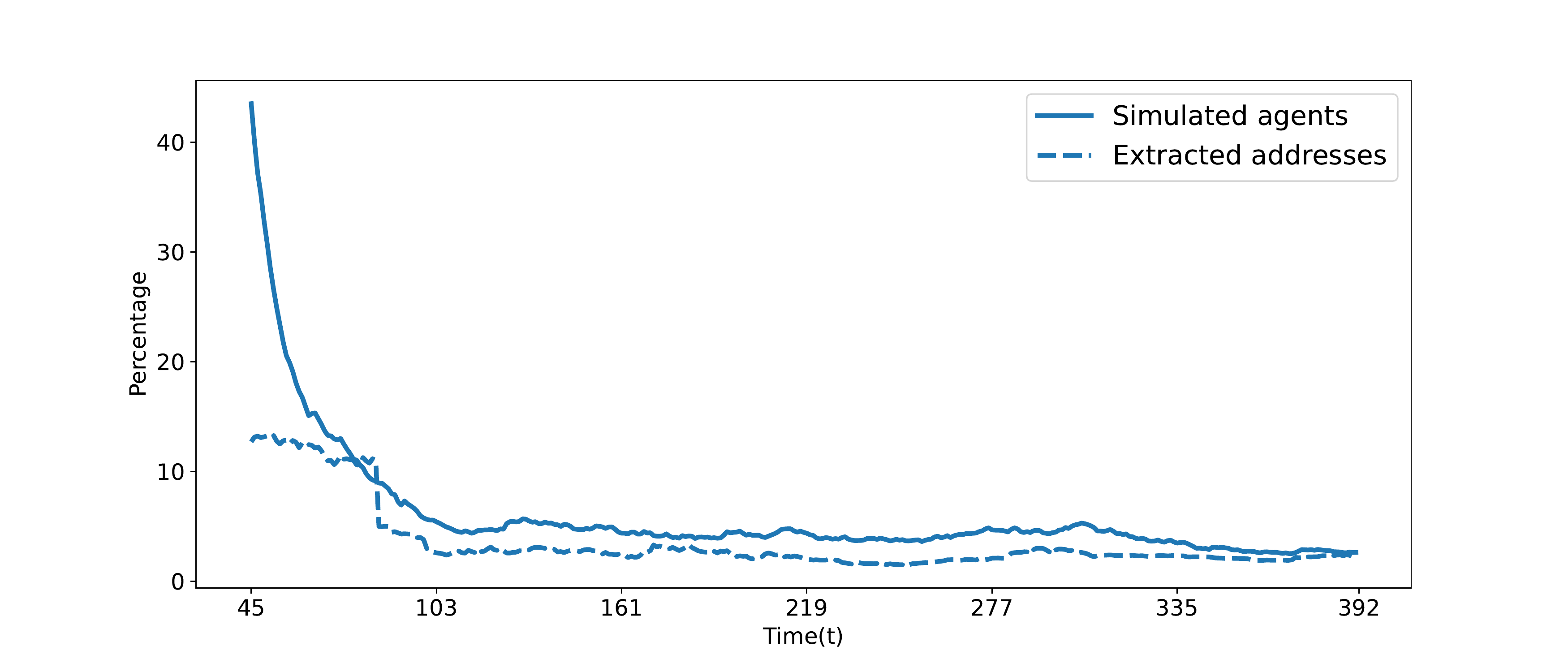}
    }
    \caption{Verification through event validity for the share of agents that control 90\% of the circulation between Yearn Finance and $S_0$.}
    \label{fig:verification_event}
\end{figure}

For the event validity, simulated events are compared with those occurring in real world systems \citep{beese2019simulation}. Specifically, we take the share of agents that control 90\% of TKNs in circulation between t=1 and t=392 for $S_0$ Cronje. (The values at t=392 are identical to those displayed in Table \ref{tab:whales}.) The real world (extracted) data is taken from Yearn Finance (c.f. Section \ref{sec:data_prep}). The comparison between these datasets is displayed in Figure \ref{fig:verification_event}.

The model during calibration was not given any information regarding the token concentration. From the Figure, we visualize how both the simulated and real world values converge after approximately 100 steps (100 natural days). At the end of the simulations the difference is 0.4 percent points. These results present a solid base for the validity of the model as it closely replicates reality \citep{davis2007developing}.

\begin{figure}[h!]
    \centering
    \makebox[0pt]{
    \includegraphics[width=0.9\columnwidth]{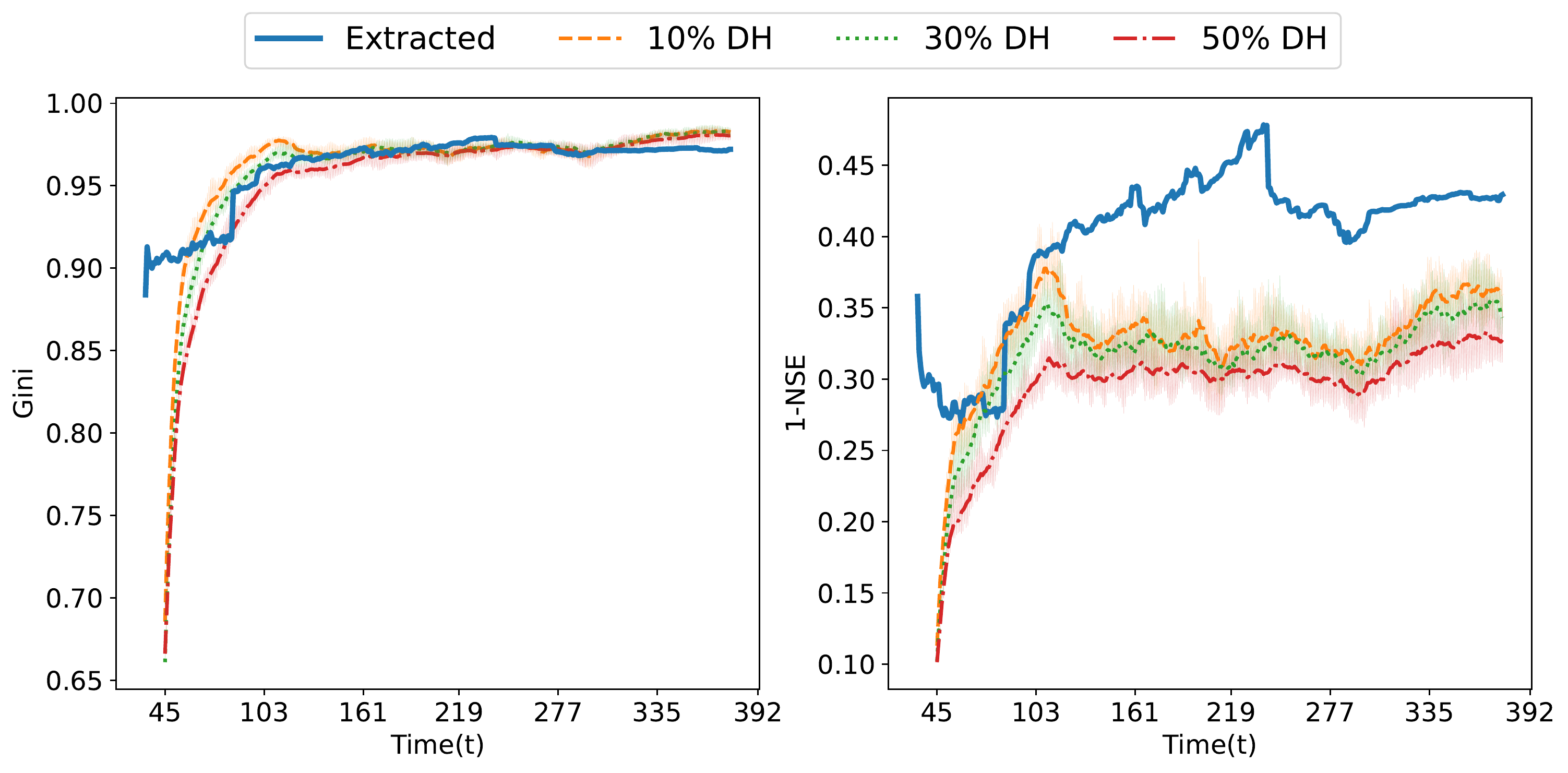}
    }
    \caption{Impact of different population allocations on NSE and Gini}
    \label{fig:verification_parameters}
\end{figure}

For the parameter variability, input parameters are modified and the resulting changes analyzed \citep{beese2019simulation}. We evaluated the impact of alternative DH/RT population ratios -- with 10\%, 30\%, and 50\% DH -- on the scenario $S_0$ relative to reality. The change was observed in terms of the metrics. Here too, we applied the Monte-Carlo method using the HPC. Figure \ref{fig:verification_parameters} gives the simulated metrics for the three ratios along the actual distribution of Yearn Finance. 

The procedure yields variability between the different population ratio and reality. We define the $\Delta$ as the difference between the simulated scenario and the extracted data. The simulations with 50\% DH ($\Delta_{Gini}$=0.71\%, $\Delta_{NSE}$=25\%) performs relatively worse than those with 10\% ($\Delta_{Gini}$=0.4\%, $\Delta_{NSE}$=16.52\%) and 30\% ($\Delta_{Gini}$=0.015\%, $\Delta_{NSE}$=19.65\%). In consideration of the $\Delta$ values and \citet{cocco2017_abm_crypto} (who take 70\% irrationality), the 30\% DH is most appropriate and therefore justifies the models' validity \citep{davis2007developing}.

%% file: sections/06_discussion.tex
\section{Discussion}
\label{sec:discussion}

%\subsection{Implications for theory} 
Using agent-based analysis, we evaluated how trading probabilities affect concentration over time within three distinct scenarios representative of 'fair' initial token allocations. Our findings are consistent with \citet{barbereau_decentralised_2022} timocratic description as the ability to trade voting rights tokens appears to be one of the causes of concentration (RQ1). Amid all three simulation sets with high, medium, and low trading probabilities, the scenarios tend towards concentration (RQ2). The concentration of wealth in the long term as observed in our constructed \ac{ABM} aligns with findings on the concentration of wealth in Bitcoin and Ethereum \citep{gochhayat_measuring_2020}, and general understandings of the concentration of wealth and inequalities \citep{piketty_capital_2014}. The implications of our findings are both of theoretical and practical nature. 

Our findings allow to ``sharpen'' theory \citep[p. 440]{davis2007developing} on tokenomics. Specifically, by contributing to the token classification of \citet{oliveira2018token} and refine the ``Governance Parameters'' in favor a distinction of the ``Supply'' parameter in terms of ``Distribution'' and ``Allocation''.
%\reviewer{Table 6 presents an excerpt from the prior work of Oliveira et al (2018). In my view, your paper needs a broader context and introduction of this token classification so that a reader can understand how your allocation dimension fits within this framework. I was also confused by the representation and incentive system dimensions (which are not introduced). How should I interpret them? Are they relevant in your context or can be ignored? This part lacks important background information from the original paper, and therefore, it is hard to objectively judge on your theoretical contribution.} \tbnote{I've added a footenote about this. We can put in the text if need be.}
The ``Distribution'' parameter is accounted for already as it is equivocally used for the ``Supply'' of tokens. For ``Allocations'' we distinguish between ``Fair Launch'' allocations (such as the ones described) and all other token allocations that may favor a minority of insiders (e.g., like Uniswap did). This contribution parallels research on \acp{ICO} which account for these 'unfair' allocations as so-called ``private pre-sale[s]'' \citep[p. 10]{fridgen2018_ICO} -- a terminology we adopt here. Table \ref{tab:token_classification_extended} showcases our refinement vis-à-vis the original classification of \citet{oliveira2018token}. While our findings do not allow to distinguish causation or correlation between allocation and concentration of tokens over time, the inclusion of ``Allocations'' in the token classification provides an indication for the normative ambitions of on-chain governance systems.

\begin{table}[ht!]
\renewcommand{\arraystretch}{2}
\caption{Italicized refinements to the Token Classification of \citet{oliveira2018token}. \label{tab:token_classification_extended}}

\centering
\resizebox{\columnwidth}{!}{%
\begin{tabular}{|l|l|c|llllllllllll} 
\hline
\multirow{4}{*}{\vspace{-0.8cm}\textbf{Governance Parameters}} & \multicolumn{2}{c|}{\textbf{Representation}}            & \multicolumn{4}{>{\centering\arraybackslash}p{4cm}|}{Digital}           & \multicolumn{4}{>{\centering\arraybackslash}p{4cm}|}{Physical}                             & \multicolumn{4}{>{\centering\arraybackslash}p{4cm}|}{Legal}                                                            \\ 
\cline{2-15}
                                       & \multirow{2}{*}{\textbf{Supply}} & \textit{\textbf{Distribution}}         & \multicolumn{3}{>{\centering\arraybackslash}m{3cm}|}{Schedule-based} & \multicolumn{3}{>{\centering\arraybackslash}m{3cm}|}{Pre-mined, scheduled distribution} & \multicolumn{3}{>{\centering\arraybackslash}m{3cm}|}{Pre-mined, one-off distribution} & \multicolumn{3}{>{\centering\arraybackslash}m{3cm}|}{Discretionary}   \\ 
\cline{3-15}
                                       &                         & \textbf{\textit{Allocation}}           & \multicolumn{6}{c|}{\textit{Fair Launch}}                                                             & \multicolumn{6}{c|}{\textit{'Unfair' launch, pre-sale}}                                      \\ 
\cline{2-15}
                                       & \multicolumn{2}{c|}{\textbf{Incentive system}}          & \multicolumn{3}{c|}{Enter Platform} & \multicolumn{3}{c|}{Use Platform}                      & \multicolumn{3}{c|}{Stay Long-Term}                  & \multicolumn{3}{c|}{Leave Platform}  \\ 
\hline
                       
\end{tabular}%

}
\end{table}

%\subsection{Implications for practice} 
The practical implications of our findings are for the design of future governance systems that leverage voting rights tokens. Our work provided additional evidence that trading largely determines the extent to which governance power is concentrated. Hence, the possibility to transfer tokens must be addressed. In practice, this can be achieved through a new class of tokens described as \textit{soulbound}. The introduced definition refers to ``accounts, or wallets, that hold publicly visible, non-transferable (but possibly revocable-by-the-issuer) tokens'' \citep[p. 2]{weyl2022decentralized}. In other words, the (albeit pseudonymous) identity of a holder is encrypted into an \ac{SBT} that is linked to the respective wallet. The opportunities for on-chain governance are promising:
\begin{itemize}
    \item They mitigate Sybil attacks.
    \item They (could) grant more voting power to reputable holders.
    \item They enable for ``proofs-of-personhood''.
    \item They allow to correlate between \acp{SBT} which support particular causes and prevent a ``tyranny of the majority'' \citep{mill1998liberty}.
\end{itemize}

These opportunities provide avenues for research as they require contextual analysis. To date, we note the application of \acp{SBT} for Know-Your-Customer processes and user credentials as the crypto-asset exchange Binance stipulated the intend to explore \ac{SBT} on its native blockchain. Notably, Binance's \ac{SBT} BNB would grant access to specific functions of the BNB Chain \citep{partz_SBT_2022}.

%% file: sections/07_conclusion.tex
\section{Conclusion}
\label{sec:conclusion}

Within the \ac{DeFi} space, recent scholarship observed the implementation of on-chain governance mechanisms for \acp{DAO}. Voting rights tokens are a key ingredient to implement these mechanisms. The initial allocation of voting rights tokens ought to follow principles of fairness in order to achieve normative goals of political decentralization. The fair launch allocation of Andre Cronje gained prominence as it did not allocate any tokens to a minority of insiders. However, in practice it fell short as over time YFI tokens became highly concentrated. On the basis of \citet{cocco2017_abm_crypto} and \citet{rocsu2021evolution}, in this paper we propose an \acf{ABM} to simulate fair launch initial token allocation. Using the model, we simulated alternative initial token allocation scenarios understood as 'fair' \citep{mill_1864,rawls1971theories}. In all of them, independently of market conditions and agents' willingness to trade, concentration looms. 

The research is subject to a number of limitations. 
The first pertains to the defined market rules. Though these followed the principles stipulated in \citet{mendelson1982market}, the clearing mechanism lacks a formal price clearing method. Additionally, despite having designed our model on the basis of \citet{cocco2017_abm_crypto}, we did not implement limit orders.
The second pertains to the awareness of agents. In the designed model the decision making of individual agent's does not depend on past decisions or those of other agents. To address these shortcomings, future work may build upon and extend our model to include a public order book where agents are aware about other orders. Further, a distinction may be made between trading mechanisms and clearing methods on centralized and decentralized exchanges.
The third pertains to the trading behavior of agents. For our simulations, we heavily rely on the FGI as a proxy for market conditions. Subsequent work could opt for the use of more granular indicators, such as the price of different \ac{DeFi} assets or social media data. 

%So what changes need to be made in order for successful  on-chain governance? A prerequisite may be to detach the tokens \tbnote{finish here}

%\begin{enumerate}
%    \item When the volatily is low and as result the whales cannot acquire a large number of tokens and then it leads to low centralization values.
%    \item Contrary when we create high vol scenario, there is enough token supply in the market, the whales aqcuire and thence the centrality increases.
%    \item trading prob related with low error rates, more trading, less error
%\end{enumerate}

%% file: sections/08_aknowledgments.tex
\section*{Acknowledgments}
\label{sec:ack}

The authors thank Reilly Smethurst for his friendly review and contributions to literature. The authors also thank Gilbert Fridgen for his valuable feedback as part of the first submission.

Joaquin Delgado Fernandez is supported by the European Union (EU) within its Horizon 2020 programme, project MDOT (Medical Device Obligations Taskforce), Grant agreement 814654. Tom Barbereau is supported by PayPal and the Luxembourg National Research Fund (FNR) -- (P17/IS/13342933/PayPal-FNR/Chair in DFS/ Gilbert Fridgen). PayPal’s financial support is administered via the FNR, and by contractual agreement, PayPal has no involvement in Tom Barbereau’s research. Orestis Papageorgiou is supported by the Luxembourg National Research Fund (FNR) (C20/IS/14783405/FIReSpARX).

%% file: sections/99_acronyms.tex
\begin{acronym}
\acro{ABM}{agent-based modeling}
\acro{DAO}{Decentralized Autonomous Organization}
\acro{dApp}{Decentralized Application}
\acro{DeFi}{Decentralized Finance}
\acro{DEX}{decentralized exchange}
\acro{ICO}{Initial Coin Offering}
\acro{ERC}{Ethereum Requests for Comment}
\acro{IS}{Information System}
\acro{DLT}{distributed ledger technology}
\acro{CI}{confidence interval}
\acro{SBT}{Soulbound Token}
\end{acronym}